\documentclass[aps,twocolumn,superscriptaddress,groupedaddress,10pt,nofootinbib]{revtex4} 

\usepackage{graphicx}  	
\usepackage{dcolumn}   	
\usepackage{bm}        	
\usepackage{amssymb}   	
\usepackage{amsmath}
\usepackage{epstopdf}

\usepackage[a4paper,margin=20mm]{geometry}
\usepackage{color}

\begin{document}

\title{Controlling Shear Jamming in Dense Suspensions via the Particle Aspect Ratio}

\author{Nicole M. James}
\affiliation{James Franck Institute, The University of Chicago, Chicago, Illinois 60637, USA}
\affiliation{Department of Chemistry, The University of Chicago, Chicago, Illinois 60637, USA}

\author{Huayue Xue}
\affiliation{James Franck Institute, The University of Chicago, Chicago, Illinois 60637, USA}
\affiliation{Department of Physics, The University of Chicago, Chicago, Illinois 60637, USA}

\author{Medha Goyal}
\affiliation{James Franck Institute, The University of Chicago, Chicago, Illinois 60637, USA}
\affiliation{Department of Physics, The University of Chicago, Chicago, Illinois 60637, USA}

\author{Heinrich M. Jaeger}
\email[E-mail:]{h-jaeger@uchicago.edu} 
\affiliation{James Franck Institute, The University of Chicago, Chicago, Illinois 60637, USA}
\affiliation{Department of Physics, The University of Chicago, Chicago, Illinois 60637, USA}

\begin{abstract}Dense suspension of particles in a liquid exhibit rich, non-Newtonian behaviors such as shear thickening  and shear jamming. 
Shear thickening is known to be enhanced by increasing the particles' frictional interactions and also by making their shape more anisotropic. For shear jamming, however, only the role of interparticle friction  has been investigated, while the effect of changing particle shape has so far not been studied systematically. 
To address this we here synthesize smooth silica particles and design the particle surface chemistry to generate strong frictional interactions such that dense, aqueous suspensions of spheres exhibit pronounced shear jamming. 
We then vary particle aspect ratio from  $\Gamma$=1 (spheres) to  $\Gamma$=11 (slender rods), and perform rheological measurements to determine the effect of particle anisotropy on the onset of shear jamming and its precursor, discontinuous shear thickening. 
Keeping the frictional interactions fixed, we find that increasing aspect ratio significantly reduces $\phi_m$, the minimum particle packing fraction  at which shear jamming can be observed, to values as low $\phi_m=33\%$ for $\Gamma$=11. 
The ability to independently control particle interactions due to friction and  shape anisotropy yields fundamental insights about the thickening and jamming capabilities of suspensions and provides a framework to rationally design shear jamming characteristics.
\end{abstract}

\maketitle


Dense particulate suspensions often display a variety of non-Newtonian flow properties, including shear thinning, shear thickening, and reversible, shear-induced solidification called shear jamming \cite{Brown2009,Peters2016}.
Shear jamming and discontinuous shear thickening (DST), where a suspension experiences a discontinuous jump in viscosity at a critical shear rate \cite{Hoffman_1972}, are both understood to depend heavily on interparticle friction \cite{Lucio_2013,Lin2015,Royer2016,Friction_2,Friction_3,Friction_1}.
The prevailing models \cite{Wyart_2014,Singh2018,Morris2018} provide a stress-dependent mechanism that drives these phenomena: 
at low applied shear, most particles in the suspension are lubricated by solvent layers, thus only weak thickening is observed.
As more particles are added the frictionless jamming packing fraction $\phi_0$ governs the onset of rigidity.
At high applied shear, a large fraction of particles are forced into close proximity, such that the lubrication layer ruptures or is reduced to molecular length-scales, at which point the continuum models describing lubrication break down. 
As a result, particles are effectively in direct contact and experience friction. 
In the presence of frictional interactions, jamming can now occur at a minimum jamming packing fraction $\phi_m$  less than $\phi_0$.
In this way, a suspension that is fluid-like at rest or low applied shear, \emph{i.e.}, has a packing fraction $\phi < \phi_0$, can be sheared into a jammed state as long as $\phi > \phi_m$. In other words, shear jamming can be observed for any $\phi$ such that $\phi_0>\phi>\phi_m$. In contrast to shear-induced aggregation \cite{Kumar2014}, the stress dependence of the frictional interactions means that shear jamming is reversible \cite{Liu2010,Waitukaitis_2012}: 
when the applied stress is removed, the suspension relaxes back to the fluid state. 

\begin{figure*}[]
    \centering
    \includegraphics[width=\linewidth]{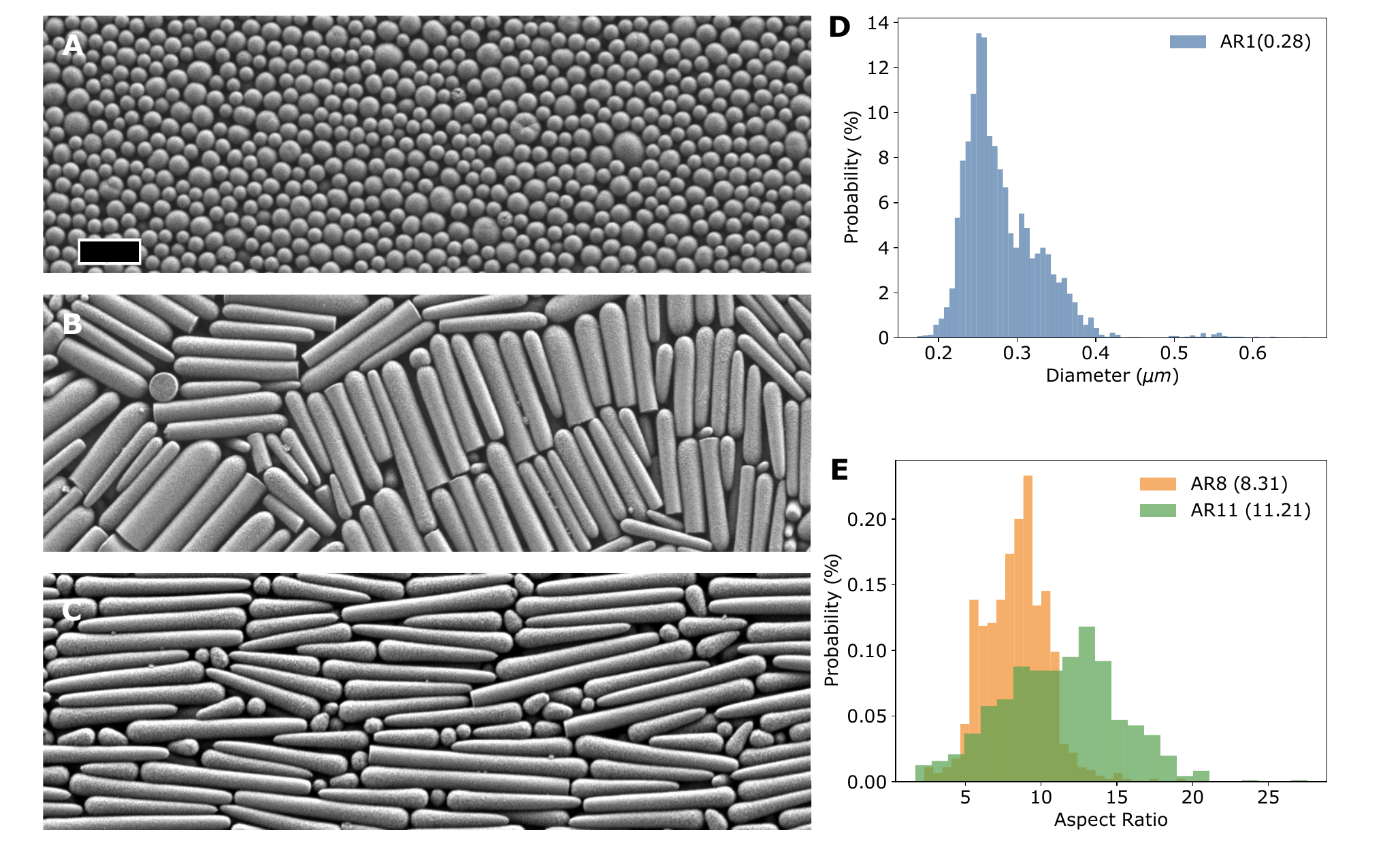}
    \caption{SEM images and characterization for AR1, AR8 and AR11 particle systems. A. AR1 spheres (scale bar indicates 1 $\mu$m), B. AR8 rods, and C. AR11 rods. The scale bar in A. applies to subplots A-C. D. Diameter distribution for AR1 spheres. E. Aspect ratio distribution for AR8 rods and AR11 rods. The mean aspect ratio for each system is stated in parentheses.}
    \label{fig:Fig1}
\end{figure*}

Shear thickening and shear jamming make dense suspensions not only challenging for industrial processing but also of particular interest for protective wear applications that rely on the material's unique dynamic response \cite{Armor,NormWagner_STF, Majumdar2013,Nenno2014,Cwalina2016}.
In all these cases, it is highly desirable to control the onset of thickening or jamming, which involves an understanding of how the key parameters $\phi_0$ and $\phi_m$ depend on particle-level properties such as interparticle friction and particle shape. 
On geometric grounds, we can expect that increasing the particles' aspect ratio by elongating them into rods will increase the number of (frictionless) contacts with neighboring particles and therefore reduce $\phi_0$. 
This is borne out by prior work on rod-shaped colloids \cite{AnisotropicDryJamming_2} and dry granular materials \cite{AnisotropicDryJamming_1}, which investigated the limit of low applied stress up to yielding or the shear thinning regime just beyond yielding.
In their sheared state, however, rod-like particles will tend to align and  \emph{a priori} it is not clear whether shear thickening will be enhanced or diminished as particle aspect ratio increases. Nevertheless, increasing aspect ratio has  been found to lower the particle concentration at which shear thickening becomes observable \cite{Egres2005,JBrown_Rods,Lutkenhaus2018}.
Still, as far as control over the shear jamming regime is concerned, the understanding of  the interplay of particle friction and  shape remains very much incomplete. 
In particular, while strong frictional interactions appear to be required in order to observe shear jamming across an appreciable packing fraction interval [$\phi_m, \phi_0$], how particle shape can affect $\phi_m$ has not been investigated in either simulations or experiments.

Here we report on experiments that aim to disentangle the effects of friction and anisotropic particle shape.  We recently demonstrated that the high degree of  friction required for suspension shear jamming can be introduced through the interparticle hydrogen bonding capacity \cite{James2018}.
This allows us to intentionally design shear jamming behavior into a model system where shape can be altered by changing the particle synthesis conditions, while fixing the surface chemistry to insure that particles retain the same frictional interactions.

We employ an emulsion-based synthesis of silica rods \cite{Kuijk_JACS}. This synthesis results in bullet-shaped rods of  $\sim$250 nm diameter and controllable length,
In aqueous solvents the silanol (Si-OH) surface groups enable strong frictional interactions via particle-particle hydrogen bonding \cite{James2018}.
Particle sizes and shapes are characterized by scanning electron microscopy and image analysis. 
Here we discuss results for three characteristic systems, each with a different aspect ratio $\Gamma$ (length-to-diameter): $\Gamma\approx$ 1 spheres (AR1), $\Gamma\approx$ 8 rods of average length $2.0$ $\mu$m (AR8), and $\Gamma\approx$ 11 rods of average length $2.8$ $\mu$m (AR11), shown in Figure \ref{fig:Fig1}. 
Above an aspect ratio of roughly 11, this synthesis produces irregularly shaped or wavey rods \cite{Kuijk_JACS}.

\begin{figure*}[]
    \centering
    \includegraphics[]{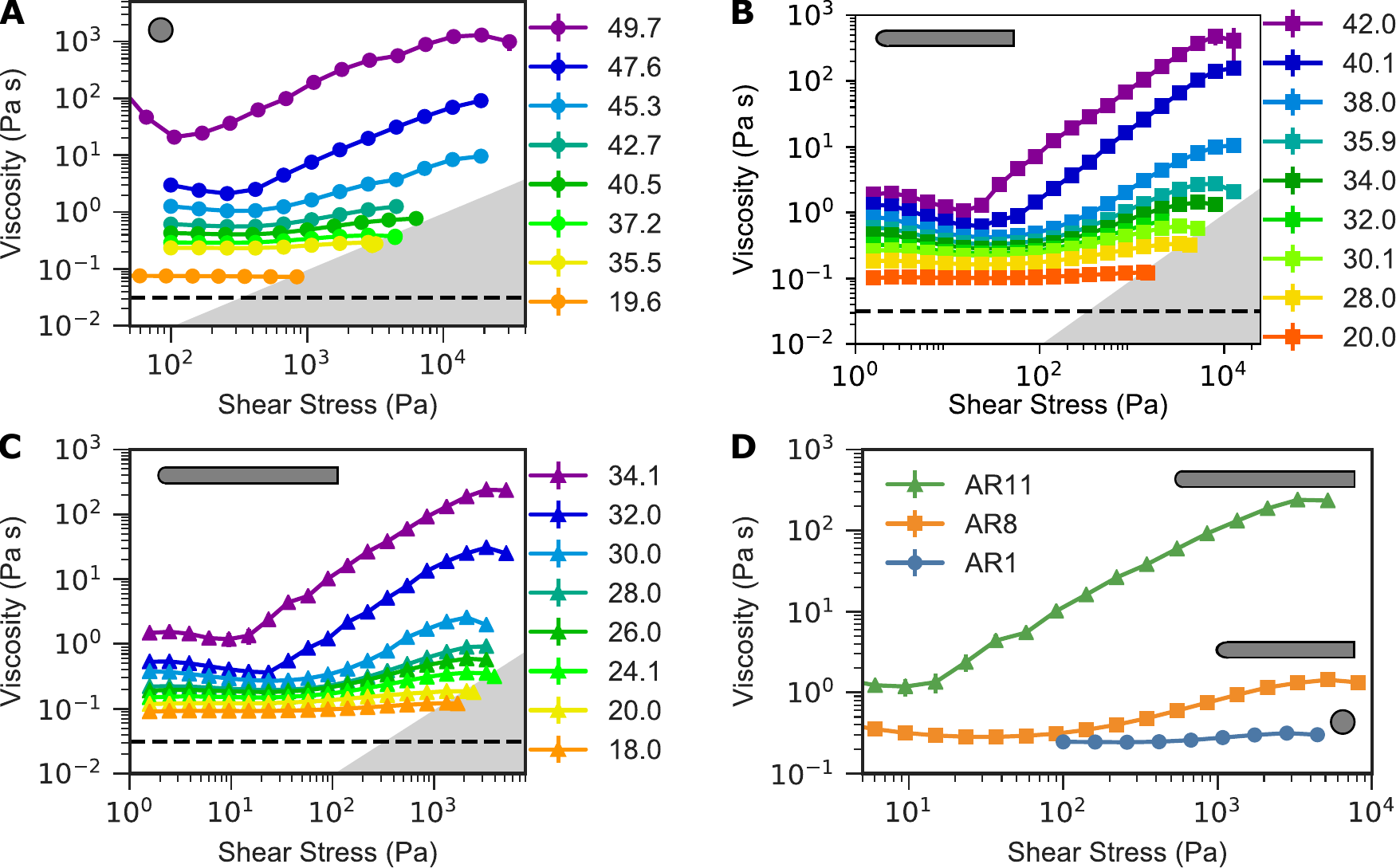}
    \caption{Rheological flow curves for A. spheres (AR1), B. aspect ratio 8 rods (AR8), and C. aspect ratio 11 rods (AR11), at various packing fractions $\phi$ (\%) in 70\% aqueous glycerol (v/v) with 15 mM NaCl. The dashed black line indicates the suspending solvent viscosity. Gray shaded regions indicate the rheometer's rate limit. Gray icons indicate the relative difference in aspect ratio (to scale) among the systems. D. Comparison of flow curves at  $\phi$=34.1 $\pm$ 0.1\%, highlighting the enhancement of shear thickening with increasing aspect ratio.}
    \label{fig:Fig2}
\end{figure*}

We perform stress-controlled rheological measurements on suspensions of these particles. 
Lack of hysteresis between increasing and decreasing stress ramps was used to determine shear protocols and ensure a steady state had been adequately established. 
In all cases, the suspending solvent was 70\% glycerol in water (v/v) with 15 mM NaCl. This solvent was selected to elicit suspension shear thickening and ensure the flow curves could be well resolved within the rheometer rate limit, indicated by the gray region in the lower-right of each plot in Figure \ref{fig:Fig2}. 
The AR1 system of spheres (Fig. \ref{fig:Fig2}a) shows mild shear thickening at $\phi= 35.5\%$ and strong, discontinuous shear thickening (DST) at $\phi=47.6\%$. 
Note that a slope of 1 on log-log plots of viscosity versus stress as in Fig. \ref{fig:Fig2} implies a vertical, discontinuous jump if the same data are plotted versus shear rate.
Concentrated packing fractions approaching $\phi=50\%$ can be prepared and remain fluid-like.
In contrast, the AR8 system (Fig. \ref{fig:Fig2}b) exhibits mild shear thickening as early as $\phi=28.0\%$, and strong DST that spans over two orders of magnitude in viscosity, at only $\phi=40.1\%$. 
Following this trend, the AR11 system (Fig. \ref{fig:Fig2}c) shows mild shear thickening at $\phi=18.0\%$, and DST at only $\phi=30.0\%$. 
To highlight the significance of this, DST with spheres typically requires packing fractions in excess of $\phi=50\%$. 

The drastic enhancement of  thickening behavior with aspect ratio at a given packing fraction is in agreement with prior studies \cite{Egres2005,JBrown_Rods,Lutkenhaus2018} 
and is shown in Figure \ref{fig:Fig2}d  by  comparing the three systems at $\phi=34.1 \pm 0.1\%$. 
While the AR1 system only very slightly shear thickens, the AR11 system already shows DST.

We now analyze the flow curves in Fig. 2 to extract the stress-dependent jamming packing fractions as a function of aspect ratio. 
To do this, we follow prior work \cite{Wyart_2014,Singh2018,Han2018} and employ a Krieger-Dougherty-type relation \cite{Krieger1959} between suspension viscosity and packing fraction

\begin{equation}
\eta_r=(1-\frac{\phi}{\phi_J})^{-\beta},
\label{eq:1}
\end{equation}

where $\eta_r$ is the suspension viscosity rescaled by the suspending solvent's viscosity, $\phi_J$ is the jamming packing fraction of interest: $\phi_0$ in the frictionless low-shear limit or $\phi_m$ in the frictional high-shear limit.
The exponent $\beta$ is a fitting parameter that is generally taken to be $\approx2$ \cite{Maron1956,Quemada1977,Singh2018,Han2018}. 

\begin{figure}[]
    \centering
    \includegraphics[]{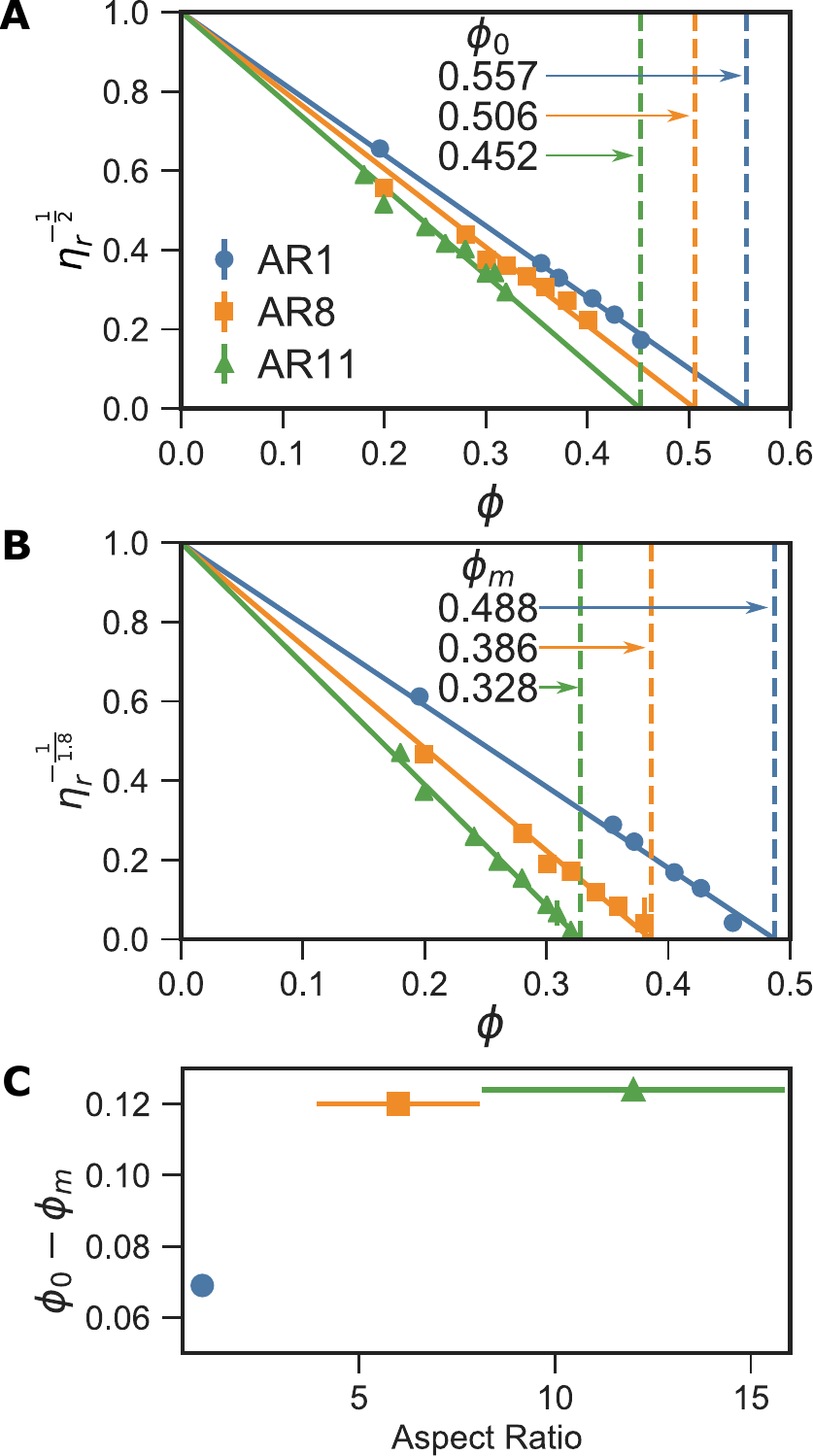}
    \caption{A. Rescaled minimum low-shear viscosities as a function of packing fraction, enabling the determination of the frictionless jamming point $\phi_0$ from linear least squares analysis. B. Rescaled maximum high-shear viscosities as a function of packing fraction, enabling the determination of the frictional jamming point $\phi_m$ from linear least squares analysis. In both A and B the exponent $\beta$ was chosen to optimize the fit for all three curves.  C. Dependence of the shear jamming packing fraction range, $\phi_0-\phi_m$, on particle aspect ratio. }
    \label{fig:Fig3}
\end{figure}

We identify  $\phi_0$ as the particle concentration at which the low-shear Newtonian viscosity (taken to be the minimum viscosity if shear thinning is present) diverges as a function of packing fraction.
From graphs as  in Fig. \ref{fig:Fig3}a, where we plot ${\eta_r}^{-1/\beta}$ as a function of $\phi$, we can read off $\phi_0$ as the intercept with the horizontal axis. 
Using  $\beta=2$ as the value that best linearized the data in such plot, this leads to $\phi_0(\text{AR1})=55.7\%$. Upon increasing the aspect ratio of the particles eight-fold, this decreases to $\phi_0(\text{AR8})=50.6\%$. 
Increasing aspect ratio further to 11 decreases $\phi_0$ again by nearly 5 points, to $\phi_0(\text{AR11})=45.2\%$. 

Similarly, we use Equation 1 to extract the frictional jamming packing fraction $\phi_m$ by considering the high-shear viscosity at the upper end of shear thickening (\ref{fig:Fig3}b), taking the maximum viscosity. In this case,  $\beta=1.8$ best linearized the data.  For some of the highest packing fractions the suspensions were so easily driven into the shear jammed state that it was experimentally unclear if shear jamming, together with slip,  occurred already just beyond the viscosity minimum (note that a solid-like, fully shear jammed state cannot be probed reliably with a steady-state viscosity measurement \cite{Hermes2016,Han2018}.
Therefore, to obtain $\phi_m$ via extrapolation we  only use data for $\phi(\text{AR1}) < 45.3\%$, $\phi(\text{AR8}) < 38.0\%$, and $\phi(\text{AR11}) < 30.0\%$. As the aspect ratio $\Gamma$ increases from 1 to 8 to 11, $\phi_m$ is found to decrease from 48.8\% to 38.6\% to 32.8\%, respectively. This large reduction by 16 percentage points in $\phi_m$ from changing particle anisotropy stands in stark contrast to the small, $\sim$1\% shift in $\phi_m$ observed when the effective, hydrogen-bonding-induced interparticle friction was changed by a factor 2 \cite{James2018}.

The packing fraction range over which shear jamming is observable is given by the interval from $\phi_m$ to $\phi_0$.
Figure \ref{fig:Fig3}c shows that this range nearly doubles as the aspect ratio $\Gamma$ is increased from 1 to 8. 
A further increase from 8 to 11 delivers only a modest enhancement.

To show these changes with aspect ratio more directly, we construct state diagrams that delineate the regimes for discontinuous shear thickening (DST) and shear jamming (SJ) as a function of packing fraction $\phi$ and applied shear stress $\tau$.  We base these diagrams on the model by Wyart and Cates \cite{Wyart_2014}, assuming infinitely hard particles and taking as input parameters the values of $\phi_m$ and $\phi_0$, obtained from Fig. 3, as well as the characteristic stress $\tau ^*$ at which the lubrication layers between particles break down and frictional interactions switch on. 
This stress value is obtained by quantitatively matching flow curves predicted by the Wyart-Cates model with data from Fig. 2, as was done in Ref. [24]. 
In these state diagrams, shown in Fig.4, $\phi_0$ is the left boundary of the regime (grey) where jamming occurs in the absence of shear simply by increasing particle density, while $\phi_m$ is the leftmost boundary of the SJ regime (green).
The DST regime is indicated by the red color.

\begin{figure}[]
    \centering
    \includegraphics[]{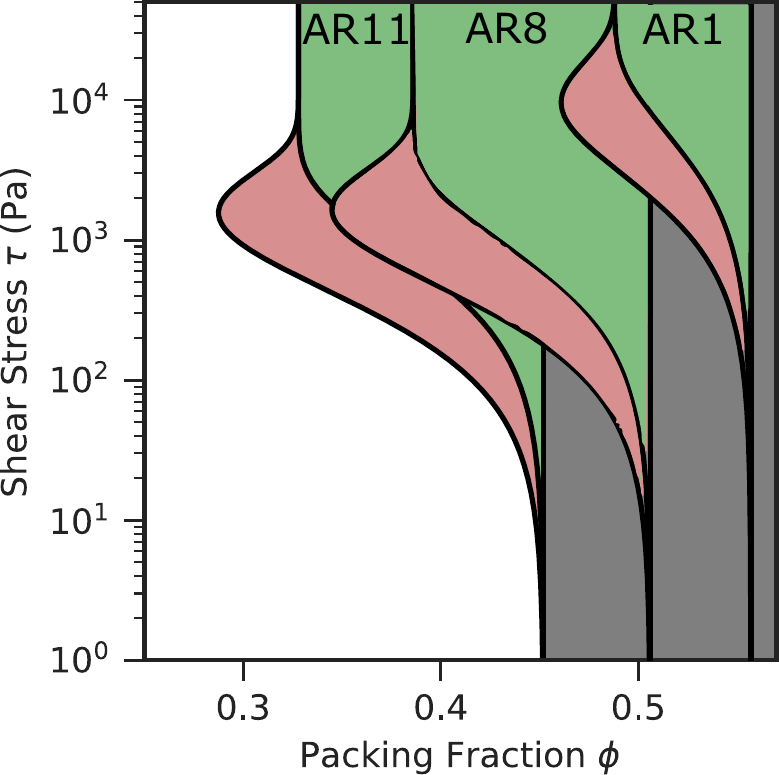}
    \caption{State diagrams for the AR11, AR8, and AR1 systems, showing regions that exhibit DST (red), shear jamming (green), or jammed-at-rest (gray) behavior.}
    \label{fig:Fig4}
\end{figure} 

In many suspensions of spherical particles that exhibit pronounced DST, an SJ regime is not readily observed \cite{James2018}. 
This is either because the frictional interactions are weak, so that $\phi_m$ is extremely close to $\phi_0$ and it becomes difficult to prepare a suspension that is not already jammed a rest, or because $\tau^*$ is so large that the stress required for SJ at any concentration $\phi$ less than $\phi_0$ becomes prohibitive. 
As Fig. 4 shows, the frictional interactions produced via the surface functionalization of the particles in this work, by contrast, generate a wide packing fraction interval for SJ already at aspect ratio $\Gamma$=1. 
Referring back to the flow curves in Fig. 2 we note that, beyond increasing the effective friction via short-range, stress-dependent hydrogen bonding, this surface functionalization does not also introduce significant longer-ranged attractive forces, which would have led to large yield stresses and a pronounced shear thinning regime \cite{Murphy2016,Lutkenhaus2018}.

The state diagrams highlight that increasing $\Gamma$ not only shifts $\phi_0$ and $\phi_m$ to lower values and enlarges the packing fraction range for DST and SJ, but also lowers the shear stress required to enter the DST and SJ regimes by almost a factor of 10.
Since all particles have the same surface functionalization this clearly demonstrates the independent role of the aspect ratio and the possibilities this opens up to control the location and extent of the DST and SJ regimes as a function of $\phi$ and $\tau$.
In fact, compared to tuning the frictional interactions for fixed spherical shape \cite{James2018}, the effect achievable by changing the aspect ratio $\Gamma$ is strikingly large.
Interestingly, most of this effect occurs up to $\Gamma$=8, while further increase of the aspect ratio does not appear to reduce the onset stress for SJ in any significant way and simply shifts the [$\phi_m, \phi_0$] interval to lower values.

These findings provide an important experimental baseline for extending current models and simulations of dense suspensions to anisotropic particle shapes.
They also provide new opportunities to design and optimize stress-adaptive materials that take advantage of DST and SJ, as used for example in protective wear or personal protective equipment that reduce injury due to impact \cite{NormWagner_STF,Armor,Cwalina2016,Majumdar2013,Nenno2014}.  
In such applications, it can be advantageous to obtain pronounced shear thickening and reversible, jamming-induced solidification with only a small amount of added particles.  
As Fig. 4 shows, with $\Gamma$=11 this can be achieved at particle concentrations as low as $\phi$ of 30$\%$ and 33$\%$, respectively.

\section*{Acknowledgements}
The authors thank Daniel Blair for helpful discussions regarding the particle synthesis, and Endao Han and Abhinendra Singh for helpful discussions. This work was supported through Army Research Office through Grant W911NF-
16-1-0078, and by the Chicago Materials Research Center/MRSEC, which is funded by the National Science Foundation through Grant NSF-DMR 1420709. HX and MG acknowledge fellowship suport through the Heising-Simons Fund.




\bibliographystyle{unsrt}
\bibliography{bibliography} 

\end{document}